# Free fall misconceptions: results of a graph based pre test of sophomore civil engineering students.

**Alicia M. Montecinos[1]**
[1]*MDCE Group, Science Faculty, Pontificia Universidad Católica de Valparaíso*, *Avenida Universidad 330 Curauma, Valparaíso, Chile.*

**E-mail:** alicia.montecinos@usm.cl

**Abstract**

A partially unusual behaviour was found among 14 sophomore students of civil engineering who took a pre test for a free fall laboratory session, in the context of a general mechanics course. An analysis contemplating mathematics models and physics models consistency was made. In all cases, the students presented evidence favoring a correct free fall acceleration model, whilst their position component versus time, and velocity component versus time graphs revealed complex misconceptions both on the physical phenomenon and it's implicit mathematics consistency. The last suggests an inability to make satisfactory connections through definitions between graphed variables. In other words, evidence strongly suggests that students are perfectly able to *memorize* the free fall acceleration model, whilst not understanding it's significance at any level. This small study originated the develope and validation of a tutorial on free fall graphs for position, velocity and acceleration models, as part of a following cross universities major project.

## I. INTRODUCTION

It is well known that free fall misconceptions are extremely common in all our students, whether they are majoring in science or not. Free fall can be addressed through energy, force or kinematics. A comprehensive paper review on the gravity force misconceptions is generously given by Kavanagh, C., and Sneider, C. [1]

A brief account of the existing literature on the kinematics perspective is given by Taşar [2], who goes deeper in something else: the mathematics perspective on position, velocity and acceleration. In his work, Taşar states that the rate of change concept is quite hard to succesfully be applied by students into a physics context, being consistently found to be a misconception among learners at various levels that is widely occurring and very resistant to change.

As Redish and Bing [3] brilliantly state: The use of physical meanings play a number of important roles in the interpretation of math in science, including: helping to guide problem solution strategies, providing metacognitive warnings to facilitate error checking, and providing reasons to reject a particular mathematical model. This reasoning can be extended to graphs. Hence, it is perfectly natural to expect that students -if asked to obtain position, velocity and acceleration graphs- would, somehow, perform an error-cheking method between them.

On a purely kinematics perspective, an open ended free response pre test results on the kinematics perpective is presented by Jugueta, Go and Indias [4], reporting expected misconceptions.

During the 2013 fall semester, a graph based pre test was applied to a small class of sophomore students. Even though position, velocity and acceleration are related through derivatives, and the students are familiar to these definitions, it was expected to find errors and misconceptions in the free fall graphs of this three variables accordingly to the available literature.

## II. METHODOLOGY

An open ended graph guided test was applied to 14 sophomore civil engineering students.

This class group had previously approved a basic calculus course and was on it's fourth week of instruction in kinematics using peer instruction strategies. Their general mechanics physics course consisted in 3 theoretical sessions (1,5 h each), 1 laboratory session (3 h) and 2 sessions with their assistant (1,5 h each).

Also, Chilean national science curriculum includes free fall motion in the second year of highschool, which has a total of 4 years, so there are 3 years in average since the last time the students discussed free fall motion.

At the beginning of the laboratory session, they were given the following instructions (Fig. 1)

**FIGURE 1**
**Free fall pre test questions**

*Part I: Suppose a small object is released from rest, falling freely in absence of air resistance. Draw the expected position component versus time graph and velocity component versus time graph, in the following space:*

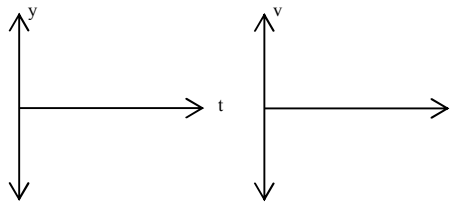

*Part II: A small object is launched vertically upwards. In absence of air, draw position, velocity and acceleration components versus time graphs. Assume that at t=T, the object reaches it maximum height.*

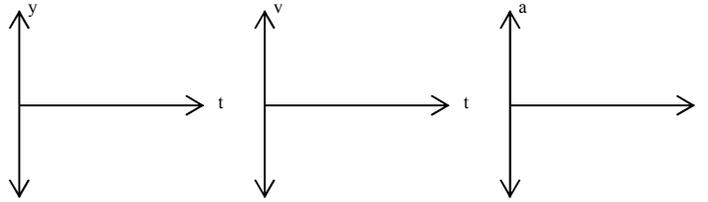

As can be seen, students had freedom to choose their reference system, in other words, the 0 m height. A cautious analysis will have to be done in order to respect their choosing.

## III. RESULTS AND DISCUSSION

*Part I*

The situation in Part I describes a **vertical downwards motion**, with zero initial velocity and constant acceleration of a body, neglecting the air resistance.

In the case of the position component versus time graph, 71% of the class draw an parabolic function, all of them acceptable models for the described situation. The rest, offered a lineal model.

In the case of the velocity component versus time graph, a more interesting behaviour was found. 86% of the class used a lineal function, from which only the 42% of the answers consisted on an acceptable model, representing the 36% of the class.

It was found that parabolic and linear models coexisted for both variables, as shown in Fig. 2:

**FIGURE 2**
**Cohexisting models for position and velocity, for a free fall starting from rest.**

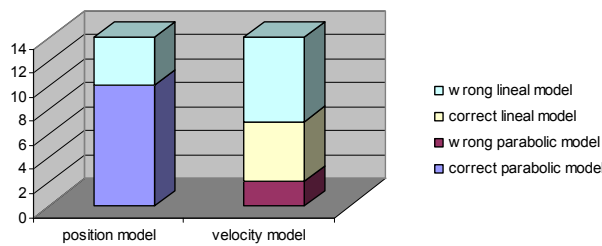

*Part II*

The situation describes the vertical motion of a body, first upwards and then downwards, with upwards initial velocity and constant acceleration, neglecting air resistance.

It was found that parabolic and lineal models coexisted only for position and velocity variables.

**For the first part of the motion**, this is, **upwards displacement,** only 57% of the class used a parabolic function for the position graph. The remaining 43% wrongly used a lineal model.

From the parabolic models, only half of them were acceptable, corresponding to the 29% of the class, as shown in Fig. 3

**FIGURE 3
Upwards motion, position versus time model. The coexisting models are practically one third each.**

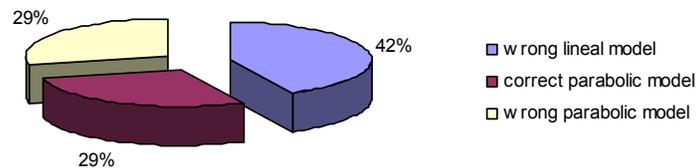

In the case of the velocity models, 79% of the class used a lineal function, whilst the rest wrongly considered it to be a parabolic one. From the lineal models, 64% of them were acceptable drawings. In other words, 50% of the class draw an acceptable lineal model for this part of the movement, as shown in Fig. 4

**FIGURE 4
Upwards motion, velocity versus time model.**

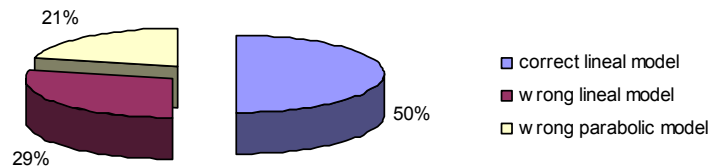

It's important to state that only 2 cases showed the traditional V-shaped speed model.

In regards of the acceleration model, 100% of the class presented an acceptable lineal model.

**For the second part of the motion**, **maximum height,** the results are the following. In the case of position, 100% of the models clearly presented a maximum height or position's scalar component, regardless of the upwards and downwards portions of the graph.

In the case of velocity, 75% of the models clearly presented a zero scalar component of velocity.

In the case of acceleration, 93% of the class had the correct model, in other words, only one student considered that at maximum height, the object had no acceleration. It is important to state that this case has correct velocity and position models for the same part of the movement.

A summary on these results are shown in Fig. 5

**FIGURE 5**
**Free fall position, velocity and acceleration models on maximum height. The variable with less mistakes, is velocity.**

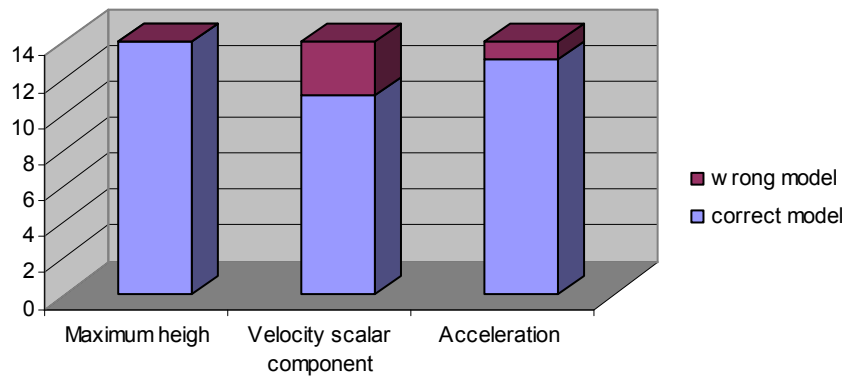

**For the last part of the motion**, this is, **downwards displacement**: 50% of the class draw a parabolic model for the position graph, from which 43% were acceptable. In other words, 21% of the class had a correct model. Details are shown in Fig. 6

**FIGURE 6**
**Downwards motion, position versus time model. Some students considered that the object remained at rest.**

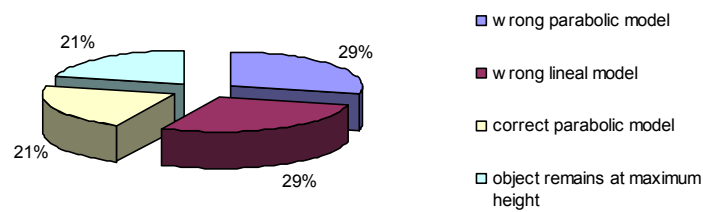

The cases in which the object did not fall back to it's starting position, presented graphs with correct constant acceleration until the end of the t-axis, and wrong position and velocity models. It is not possible to address the position mistake as a failure in reading comprehension. It is a misconception on the free fall model itself.

In the case of velocity, 75% of the class used a lineal function, whilst the rest considered it to be a parabolic model. From the lineal cases, 27% were acceptable. In other words, 21% of the class draw a correct model, as shown in Fig. 7

**FIGURE 7**
**Downwards motion, velocity versus time model. Most students correctly identofied the velocity model as a linear one, but made many different versions of it.**

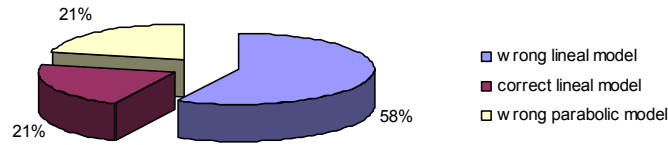

And, again, 100% of the class draw a correct acceleration model.

If considering the internal consistency between the three graphs for every student, **only the 14% of the class succeed in drawing all expected models**, showing a decrease of 60% compared to part I.

**Comparison between part I and part II/downwards motion**: for the position model, there was a decrease of 30% on the frequency of parabolic models, and a decrease of 70% correct answers. In the case of velocity, a decrease of 75% in lineal models is found, with a decrease of 40% on correct answers. This results show that the student perceive this motions as different cases, being that they have the exact same qualities, which should have been seen in the models. A comparison of only correct models frequency is made in Fig. 8:

**FIGURE 8**
**Comparison between part I and part II/downwards motion, of correct models. A decrease in the correct models frequency is found, revealing these motions are perceived as different or unrelated cases by students.**

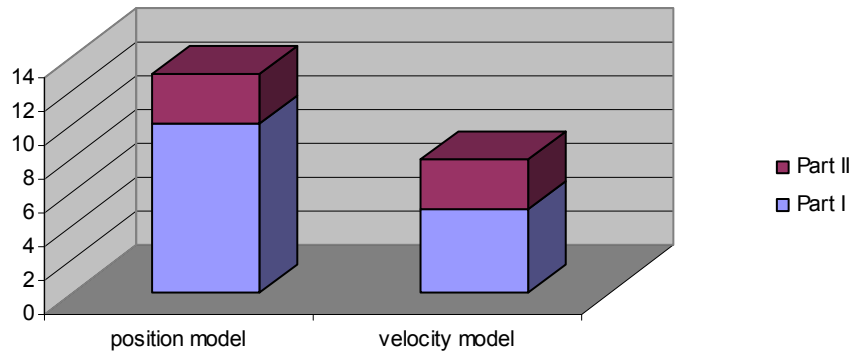

To close this analysis, a summing up pie chart is shown, considering the complete motion of part II (See Fig. 9)

**FIGURE 9**
**Summing up of the results, part II, complete motion.**

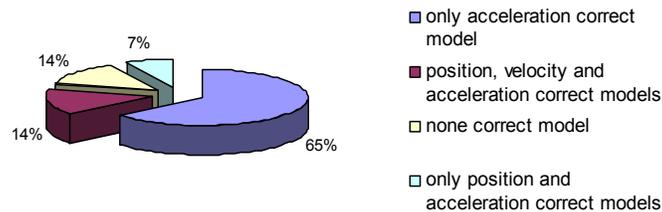

The acceleration model results are very different from the reported by Kavanagh et al, which showed more complex cases. It is proposed that this difference is due to the four weeks on kinematics instruction of the sophomore students.

If a student is sure about the acceleration of the free fall motion, so sure that he or she draws it *even though it has absolutely nothing to do* with the previos velocity model, why is he or she unable to check their relationship through a simple inspection: integers and derivatives?

## IV. CONCLUSIONS

An open-ended, graph guided pre-test was used to determine the conceptual understanding of sophomore civil engineering students to gather their previous concepts about the model of free fall, finding interesting results and further questionings.

Evidence strongly suggests that students are perfectly able to *memorize* the free fall acceleration model, whilst not understanding it's definition and significance at any level. This argument rests on the evidence of wrong position and velocity graphs.

The above can be considered as a failure in achieving a connection between these three models, through the definitions of the variables involved, as literature reports.

The main difference with the literature, though, is the extremely high percentage of students able to correctly evoke *g*, the free fall acceleration model. Another difference, is the virtual absence of zero-velocity/zero-acceleration misconception, reported by literature.

The misconceptions found in this work may be used to create multiple choice questions for assessments regarding position, velocity and acceleration for a free fall case.

No mathematical consistency check evidence was found among the results, but for 2 cases.

A good suggestion is, if a pre test like this is taken, to consider the internal consistency between the position, velocity and acceleration models, in order to evaluate the misconceptions state in regards of the connection by definition that should be clearly seen between the three graphs.

The above is based on the comparison between the individual graphs analysis results and the internal consistency analysis results, always with lower performance, thus exposing deeper trouble.

A major project of a tutorial on free fall graphs is under validation process, using the derivatives knowledge of the students as control questions (metacognition theory based and cooperative active learning approached), expected to be shared with the comunity during the first semester of 2014.

**ACKNOWLEDGEMENTS**

The author kindly regards the Civil Engineering School, Engineering Faculty, Pontificia Universidad Católica de Valparaíso, for their constant support.